# Transient Heat Conduction on Quantum Hardware: Replicated Temperature-Field Reconstruction and Shot-Count Sensitivity with a Variational Quantum Linear Solver

**Yijin Mao, Yuwen Zhang**
Department of Mechanical and Aerospace Engineering
University of Missouri
Columbia, MO 65211, USA

**Abstract**

Implicit discretizations of transient heat conduction repeatedly generate sparse linear systems, making diffusion problems a natural test bed for quantum linear solvers. This study evaluates how accurately a variational quantum linear solver (VQLS) recovers the temperature field produced by one backward-Euler step of one-dimensional conduction on IBM quantum hardware. A pre-specified confirmatory experiment compares N=4 interior temperatures encoded on two qubits with N=8 temperatures encoded on three qubits using ten paired hardware blocks; a separately defined N=16, four-qubit extension is evaluated in five independently scheduled exploratory blocks. Local VQLS optimization reduces relative temperature-field error to approximately $10^{-8}$ for all three cases, establishing an essentially exact ideal reference at this scale. On hardware, mean relative $L_2$ temperature-field error is 8.07% for N=4 and 20.14% for N=8, a paired increase of 12.06 percentage points (95% CI, 10.71-13.42; exact paired sign-flip p=0.001953). The exploratory N=16 mean is 24.75±1.84%. A pre-specified N=16 shot-count follow-up then evaluates the same locked circuit at 2,048, 8,192, and 32,768 shots in six independently scheduled blocks. Mean field error remains 23.64%, 24.57%, and 24.37%, respectively; the 32,768-minus-2,048-shot contrast is +0.74 percentage points (95% CI, -0.65 to 2.12; p=0.281). In contrast, ideal multinomial sampling predicts a reduction from 2.68% to 0.75%. Higher shot counts reduce block-to-block variability but do not recover the mean cold-side temperature field. The results show that full-field readout of a strongly nonuniform thermal state is limited by more than ordinary finite-shot sampling under the tested hardware conditions, motivating observable-focused quantum formulations for heat flux, stored energy, and selected temperatures.

**Keywords:** transient heat conduction; numerical heat transfer; variational quantum linear solver; temperature-field reconstruction; shot-count sensitivity; amplitude encoding; IBM Quantum

## 1. Introduction

Transient conduction is commonly advanced with implicit finite-difference, finite-volume, or finite-element formulations because diffusion-dominated simulations often require repeated solution of sparse algebraic systems [1-4]. The same kernel appears in more complicated thermal models, including thermal-energy-storage calculations in which conduction is coupled to phase change, convection, or temperature-dependent material properties [5-8]. In these simulations the linear-system solution is not an end in itself: engineering use depends on the temperature field and on heat fluxes, stored energy, interface conditions, and thermal loads derived from that field. Quantum linear-system algorithms provide an alternative representation of this algebraic task. The Harrow-Hassidim-Lloyd algorithm established the canonical framework [9], and later work improved precision dependence, introduced preconditioning, and extended quantum linear algebra to Poisson, finite-element, and partial-differential-equation formulations [10-18]. Practical use, however, must account for state preparation, conditioning, circuit depth, measurement cost, and extraction of classical information [19]. Variational quantum algorithms offer a hybrid route for present noisy devices [20-27]. In VQLS, a parameterized circuit prepares a candidate state while a classical optimizer minimizes a cost that measures satisfaction of the linear system [28, 29]. Related approaches have been applied to heat conduction, finite-element systems, general linear PDEs, and nonlinear differential equations [30-36].

For heat-transfer applications, full-field readout deserves particular attention. Direct amplitude encoding represents N unknowns using $\log_2(N)$ qubits, but the encoded quantum state is not yet the temperature vector required by a conventional thermal analysis. Recovering every nodal temperature can become difficult when the field spans a large dynamic range because cold-side temperatures correspond to very small basis-state probabilities. Fourier heat flux further depends on spatial temperature gradients [1], so a low global variational cost does not by itself establish that the recovered field is thermally useful. The present work therefore asks a practical question: how faithfully can current quantum hardware reconstruct an implicit transient-conduction temperature field, and can additional measurement shots recover low-amplitude thermal degrees of freedom? The study combines classical and statevector verification, a ten-block paired N=4/N=8 confirmatory experiment, a five-block exploratory N=16 extension, and a six-block N=16 shot-count follow-up. Its main contributions are a transparent conduction-to-VQLS mapping, direct hardware temperature-profile reconstruction, statistically valid block-level comparison, explicit separation of confirmatory and exploratory evidence, and a controlled test of finite-shot sensitivity using an unchanged locked N=16 circuit.

## 2. Transient Heat-Conduction Benchmark

### 2.1 Governing equation and boundary conditions

The dimensionless one-dimensional heat equation is

$$\partial T/\partial t = \alpha\, \partial^2 T/\partial x^2, \quad 0 < x < L,$$

with constant Dirichlet boundary conditions

$$T(0,t) = T_{left}, \quad T(L,t) = T_{right},$$

and a uniform initial condition $T(x,0)=T_0$. The benchmark uses $L=1$, $\alpha=1$, $T_{left}=1$, $T_{right}=0$, and $T_0=0$. These values define a canonical diffusion step from a heated boundary into an initially cold domain and keep the analysis dimensionless. A uniform grid contains N interior unknown temperatures with

$$\Delta x = L/(N+1).$$

Backward Euler in time and a centered second-order difference in space give

$$(T_i^{n+1} - T_i^{n})/\Delta t = \alpha(T_{i-1}^{n+1} - 2T_i^{n+1} + T_{i+1}^{n+1})/\Delta x^2.$$

Defining the mesh diffusion number (Fourier number)

$$r = \alpha \Delta t/\Delta x^2,$$

the implicit update becomes

$$-rT_{i-1}^{n+1} + (1+2r)T_i^{n+1} - rT_{i+1}^{n+1} = T_i^{n}.$$

Thus, each time step requires solution of

$$A\, T^{n+1} = b^{n},$$

where A is a real, symmetric, positive-definite tridiagonal matrix with diagonal entries 1+2r and first off-diagonal entries -r. Boundary contributions are included in $b^n$.

The benchmark fixes r=1. For the first time step from the zero interior field, $b=[1,0,...,0]^T$ after the left-boundary contribution is included. Holding r fixed as N changes is intentional: N=4, N=8, and N=16 are operator-size benchmarks constructed from the same nondimensional conduction stencil, not a same-time spatial-convergence study. The corresponding time increments are $\Delta t=0.0400$ for N=4, $\Delta t=0.01235$ for N=8, and $\Delta t=0.003460$ for N=16.

## 2.2 Classical reference solutions

The first-step interior reference solutions are

$$T_{ref,N=4} = [0.381818, 0.145455, 0.054545, 0.018182]^T,$$

for N=8,

$$T_{ref,N=8} = [0.381966, 0.145898, 0.055728, 0.021285, 0.008127, 0.003096, 0.001161, 0.000387]^T.$$

and for N=16,

$$T_{ref,N=16} = [0.381966, 0.145898, 0.055728, 0.021286, 0.008131, 0.003106, 0.001186, 0.000453, 1.731\times10^{-4}, 6.611\times10^{-5}, 2.525\times10^{-5}, 9.644\times10^{-6}, 3.682\times10^{-6}, 1.403\times10^{-6}, 5.260\times10^{-7}, 1.753\times10^{-7}]^T.$$

The N=8 profile spans nearly three orders of magnitude across the interior nodes, while the N=16 profile spans more than six orders of magnitude. The two-norm condition number of A increases only modestly, from 3.34 for N=4 to 4.35 for N=8 and 4.80 for N=16. Thus, the larger hardware errors observed later cannot be interpreted as a conditioning effect alone.

Table 1. Numerical, variational, and locked hardware characteristics of the three heat-conduction benchmark configurations.

| **Quantity** | **N=4** | **N=8** | **N=16*** |
|---|---|---|---|
| Interior unknowns, N | 4 | 8 | 16 |
| Grid spacing, $\Delta x$ | 0.2000 | 0.1111 | 0.05882 |
| Time increment at r=1, $\Delta t$ | 0.0400 | 0.01235 | 0.003460 |
| Condition number, $\kappa_2(A)$ | 3.342 | 4.354 | 4.802 |
| Logical qubits | 2 | 3 | 4 |
| Ansatz repetitions | 2 | 3 | 3 |
| Variational parameters | 6 | 12 | 16 |
| Sampler depth / two-qubit gates | 13 / 2 | 30 / 6 | 32 / 9 |
| Estimator depth / two-qubit gates | 12 / 2 | 29 / 6 | 31 / 9 |
| Nonzero Pauli terms, A / H / G | 4 / 9 / 6 | 8 / 21 / 14 | 16 / 45 / 30 |

*Note: *N=16 is the separately designated exploratory extension; its hardware layout differs from the shared N=4/N=8 confirmatory support.*

## 3. Variational Quantum Linear Solver

### 3.1 Cost function

The right-hand side is normalized as $|\hat{b}\rangle = b/\|b\|_2$. For the real system considered here, the VQLS cost is

$$C(\theta) = \langle\psi(\theta)|H|\psi(\theta)\rangle \,/\, \langle\psi(\theta)|G|\psi(\theta)\rangle,$$

where

$$H = A^T(I - |\hat{b}\rangle\langle\hat{b}|)A, \quad G = A^T A.$$

The cost vanishes when $A|\psi\rangle$ is parallel to $|b\rangle$, so the cost can be evaluated using expectation values of Hermitian operators. The matrices A, H, and G are decomposed into Pauli operators for execution with the Estimator primitive. The number of nonzero Pauli terms increases from 4/9/6 for A/H/G in the N=4 case to 8/21/14 for N=8 and 16/45/30 for N=16.

### 3.2 Ansatz and local optimization

A real-amplitude hardware-efficient ansatz is used. Each layer applies one Ry rotation to every logical qubit, with adjacent layers connected by a nearest-neighbor entangling chain. N=4 uses two logical qubits, two entangling repetitions, and six variational parameters; N=8 uses three qubits, three repetitions, and twelve parameters; and N=16 uses four qubits, three repetitions, and sixteen parameters. This shallow structure follows the broader hardware-efficient variational strategy used on near-term devices [18-21]. All variational parameters are optimized locally before hardware execution. COBYLA [37] minimizes the exact statevector cost, which isolates the subsequent hardware experiment from stochastic optimizer behavior and avoids expending QPU allocation on iterative parameter search. Once a normalized state $|\psi\rangle$ is obtained, its physical scale is recovered by minimizing $||A(s\psi)-b||_2$, yielding

$$s^* = (A\psi)^T b \,/\, ||A\psi||_2^2.$$

The physical temperature vector is then $Tq = s^*\psi$.

### 3.3 Reconstruction from Sampler measurements

The selected diffusion state is nonnegative. This property permits a problem-specific reconstruction from Sampler probabilities $p_i$ using

$$\psi_i = +\sqrt{p_i}.$$

The positive sign is a physical prior supplied by this benchmark; it is not general quantum-state tomography and cannot recover arbitrary signs or phases. After the amplitudes are obtained, the least-squares scaling above maps the normalized state back to the temperature vector. This reconstruction step is central to the heat-transfer interpretation. For an amplitude-encoded thermal state, $p_i$ is proportional to $T_i^2$ after normalization, so small temperatures are represented by quadratically smaller probabilities. A field that decays steeply toward a cold boundary can therefore be straightforward to encode while remaining difficult to reconstruct from finite measurements. Increasing the number of shots reduces multinomial sampling variability, but it cannot by itself correct systematic gate, readout, or state-preparation errors; the N=16 follow-up described below was designed to separate these effects empirically.

## 4. Hardware Experiment and Statistical Design

### 4.1 Backend, compilation, and locked circuits

All reported hardware measurements were acquired on IBM ibm_miami [38] using Qiskit 2.5.2 and qiskit-ibm-runtime 0.47.0 [39]. The transpiler optimization level was 2 and the transpiler seed was fixed. For the confirmatory N=4/N=8 experiment, a connected physical path [9,19,18] was shared across benchmark sizes: N=4 occupied [9,19], while N=8 used [9,19,18]. The locked Sampler depths were 13 and 30, the Estimator depths were 12 and 29, and the two-qubit-gate counts were 2 and 6 for N=4 and N=8, respectively. The later N=16 extension was independently locked on connected physical support [62,63,64,74], with Sampler depth 32, Estimator depth 31, and nine two-qubit gates in each circuit. Exact instruction-set-architecture circuits were reproduced from the backend target at the lock time. Differences in layout, circuit structure, operator decomposition, and calibration window are therefore part of the benchmark configurations and are not treated as independently controlled causal variables.

### 4.2 Confirmatory paired hardware blocks and Runtime primitives

Ten paired hardware blocks were acquired. Each block contained matched N=4 and N=8 evaluations for both Sampler and resilience-0 Estimator measurements, so each block contributed one paired observation for each endpoint. In Qiskit Runtime terminology, a Primitive Unified Bloc (PUB) is the primitive input unit for a circuit together with its associated parameters or observables. Each paired Sampler job contained one N=4 PUB and one N=8 PUB at 2048 shots per PUB; each paired Estimator job contained one N=4 PUB and one N=8 PUB at 1024 shots per PUB. The block, rather than a shot or PUB, was treated as the experimental unit. This distinction prevents finite-shot technical replication from being mistaken for independent hardware evidence. Case order was balanced across the ten blocks. Runtime scheduling metadata were retained for reproducibility but were not used as inferential factors.

### 4.3 Exploratory N=16 extension

After completion of the pre-specified N=4/N=8 confirmatory experiment, a four-qubit N=16 extension was defined to probe full-field readout at a larger encoded dimension. Five independently scheduled hardware blocks (E01-E05) were acquired using the locked [62,63,64,74] support. Each block contained one 2,048-shot Sampler evaluation and one 1,024-shot resilience-0 Estimator evaluation submitted within the same IBM Batch. Blocks were separated in execution time and each block was treated as one exploratory hardware replicate. The N=16 extension was not added to the confirmatory N=4/N=8 hypothesis test; its field error, residual, VQLS cost, and nodewise reconstructed temperatures are summarized descriptively across the five block-level observations.

### 4.4 N=16 shot-count sensitivity follow-up

To test whether the N=16 reconstruction error was dominated by finite sampling, a mechanistic follow-up was specified before any shot-sensitivity hardware data were acquired. The immutable N=16 Sampler circuit, optimized variational parameters, physical support [62,63,64,74], transpiler seed, and reconstruction procedure were retained. Six independently scheduled blocks (S01-S06) each contained Sampler evaluations at 2,048, 8,192, and 32,768 shots. All six possible within-block shot-order permutations were used exactly once. Each shot condition was submitted as a separate Runtime job within the same IBM Batch, and the independently scheduled block - not a shot or job - was treated as the experimental unit. Estimator measurements were not repeated because the follow-up targeted readout of the fixed prepared state.

A zero-QPU control quantified the finite-shot limit expected from ideal sampling. For each shot level, 10,000 multinomial draws were generated from the ideal N=16 probability distribution using a fixed random seed. These Monte Carlo draws are model-based sampling realizations rather than independent hardware replicates; they provide a reference for the error and uncertainty attributable to finite sampling alone.

### 4.5 Endpoints and statistical analysis

The pre-specified primary endpoint was the paired difference in relative temperature-field error,

$$d_j = \varepsilon_{T,N=8,j} - \varepsilon_{T,N=4,j}, \quad j=1,\ldots,10,$$

where

$$\varepsilon_T = ||T_q - T_{ref}||_2 \,/\, ||T_{ref}||_2.$$

For the confirmatory N=4/N=8 study, the mean paired difference in relative temperature-field error is reported with a 95% t confidence interval and a two-sided exact paired sign-flip randomization p-value. Secondary confirmatory endpoints are the paired relative linear-system residual and resilience-0 Estimator VQLS cost; Holm adjustment is applied across the two secondary p-values. Confirmatory inference was released only after all ten paired blocks were complete. The N=16 extension is reported separately using the sample mean, sample standard deviation, and observed range across five exploratory blocks. For the shot-count follow-up, the pre-specified primary contrast is the within-block difference in relative $L_2$ field error between 32,768 and 2,048 shots, defined as higher-shots minus lower-shots so that a negative value denotes improvement. The mean paired contrast is reported with a 95% t confidence interval and a two-sided exact sign-flip p-value. Secondary contrasts are 8,192 versus 2,048 shots for field error and 32,768 versus 2,048 shots for residual, with Holm adjustment across these two secondary tests. No shot-count inference was released until all six pre-specified blocks were complete.

## 5. Results

### 5.1 Classical and ideal-quantum verification

The locally optimized VQLS solution reproduces the classical temperature vector to essentially numerical precision. Relative $L_2$ temperature-field error is $2.55\times10^{-8}$ for N=4, $4.02\times10^{-8}$ for N=8, and $6.30\times10^{-8}$ for N=16; the corresponding relative residuals are $1.65\times10^{-8}$, $2.35\times10^{-8}$, and $3.70\times10^{-8}$. For N=16, the optimized VQLS cost is approximately $1.29\times10^{-15}$. Independent NumPy and Qiskit statevector evaluations agree to numerical precision. These checks show that, for the tested dimensions, the conduction discretization, basis ordering, ansatz, and ideal variational formulation are not the dominant sources of the hardware errors discussed below.

### 5.2 Reconstructed temperature profiles

Figure 1 compares the classical first-step temperature profiles with the mean fields reconstructed from IBM hardware. N=4 and N=8 summarize the ten paired confirmatory blocks, whereas N=16 summarizes the five independent exploratory blocks. In all cases the hot-side temperatures are recovered more faithfully than the cold-side field. For N=4, the first interior mean is 0.3829±0.0020 versus the classical value 0.3818, while the discrepancy grows toward the cold boundary. For N=8, the first two interior nodes remain close to the reference, but the cold half develops a positive, non-monotonic floor. The effect is more extensive for N=16: after the first few nodes, the classical profile rapidly approaches zero while the hardware reconstruction remains at approximately 0.01-0.05 over much of the domain. Prescribed boundary temperatures are included only to complete the physical profile and carry no measurement uncertainty.

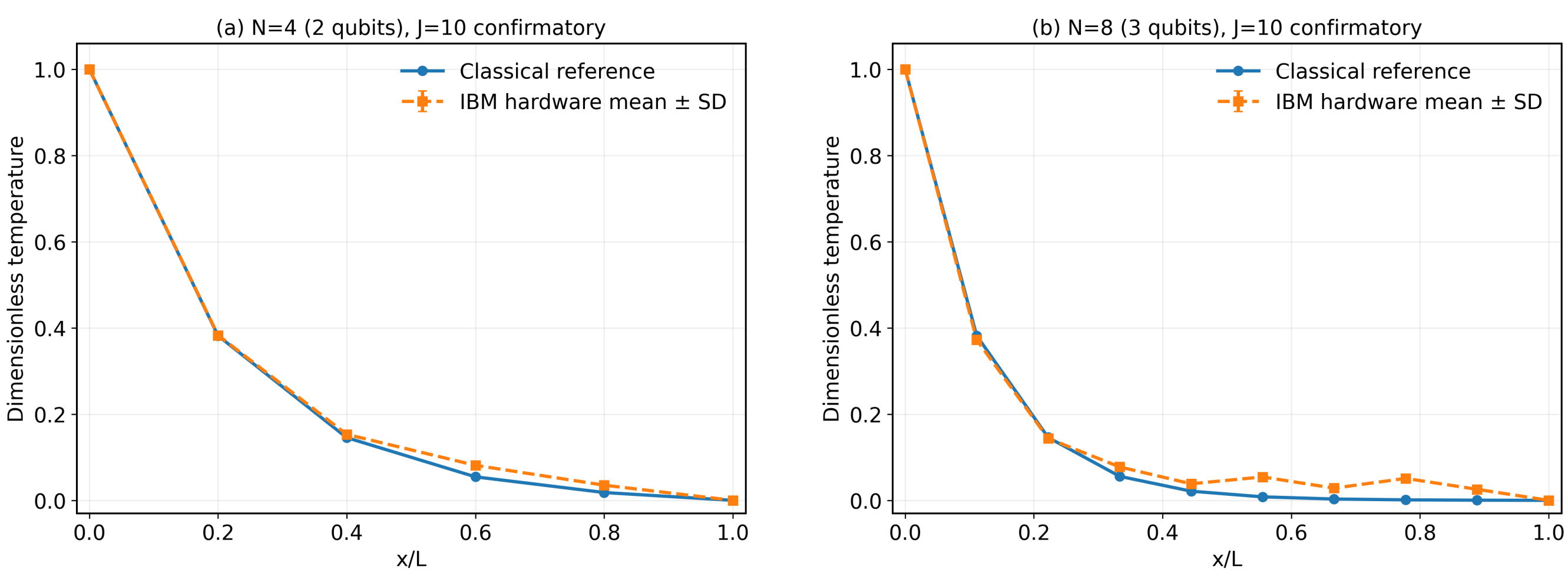

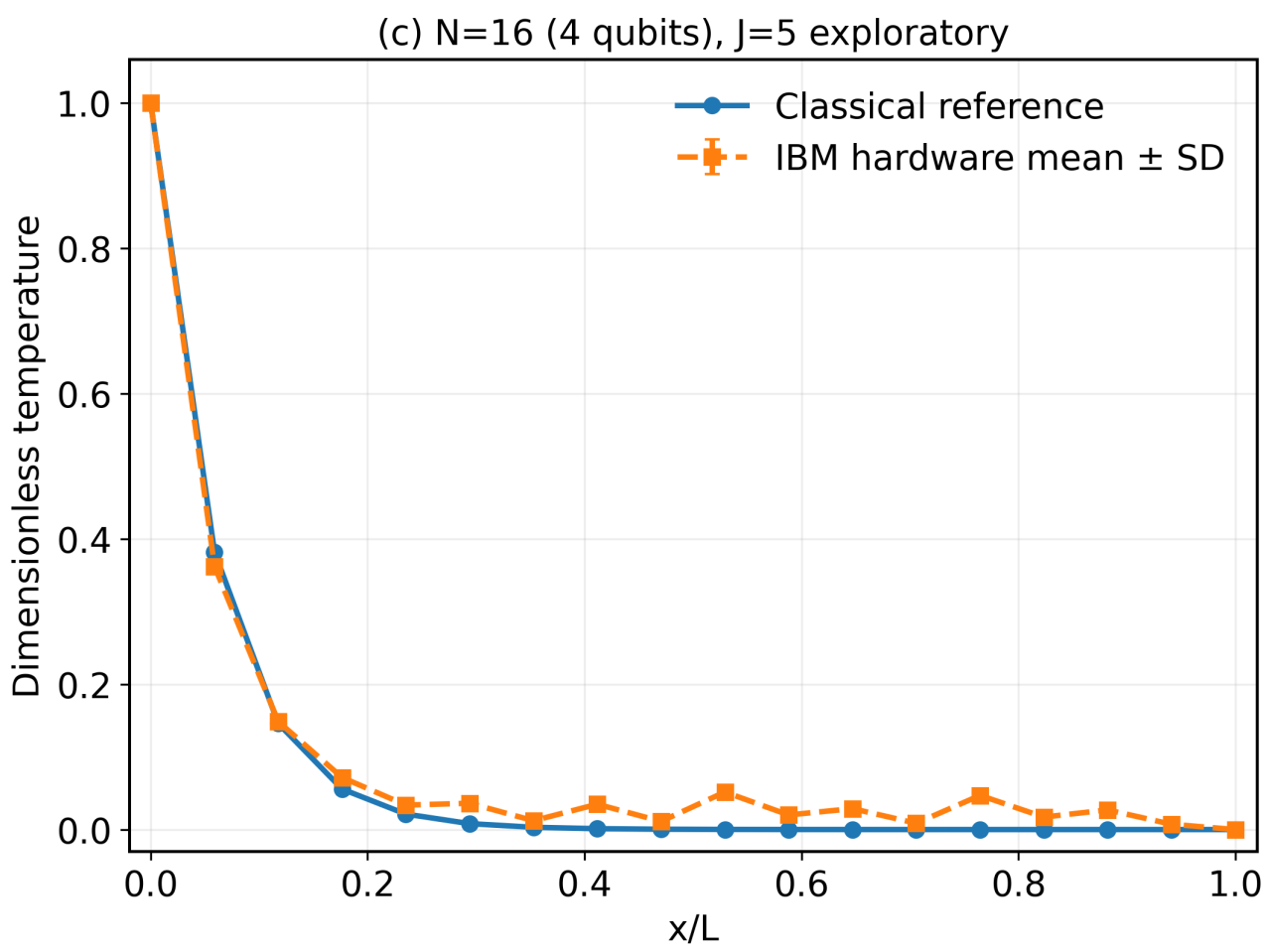

Figure 1. Classical reference and IBM-hardware reconstructed temperature profiles for the N=4, N=8, and N=16 transient-conduction benchmarks. N=4 and N=8 markers show the mean across J=10 independent confirmatory paired blocks; N=16 markers show the mean across J=5 independently scheduled exploratory blocks. Error bars denote one sample standard deviation at reconstructed interior nodes. The prescribed Dirichlet boundary temperatures are included for physical completeness and are not treated as quantum measurements.

The cold-side distortion is consistent with the probability structure of amplitude encoding. For the ideal N=8 state, the last four interior populations correspond to approximately 0.79, 0.11, 0.016, and 0.0018 expected counts at 2,048 shots. For N=16, interior nodes 5 through 16 all correspond to fewer than one expected ideal count, with the expected count falling from approximately 0.79 at node 5 to less than $10^{-6}$ near the cold boundary. Thus a large fraction of the encoded field is below the single-count scale even before hardware imperfections are considered.

### 5.3 Primary temperature-field error

The block-level relative $L_2$ error is shown in Figure 2. The N=4 mean is 8.07% with a sample standard deviation of 0.84 percentage points, whereas the N=8 mean is 20.14% with a sample standard deviation of 2.11 percentage points. The mean paired increase is 12.06 percentage points (95% CI, 10.71-13.42). All ten block-level differences are positive, and the two-sided exact paired sign-flip test gives p=0.001953. The result is therefore consistent across hardware occasions rather than being driven by a single unfavorable N=8 execution. Across the ten blocks, N=8 field error ranges from 18.4% to 25.4%, but no block reverses the N=8>N=4 relationship.

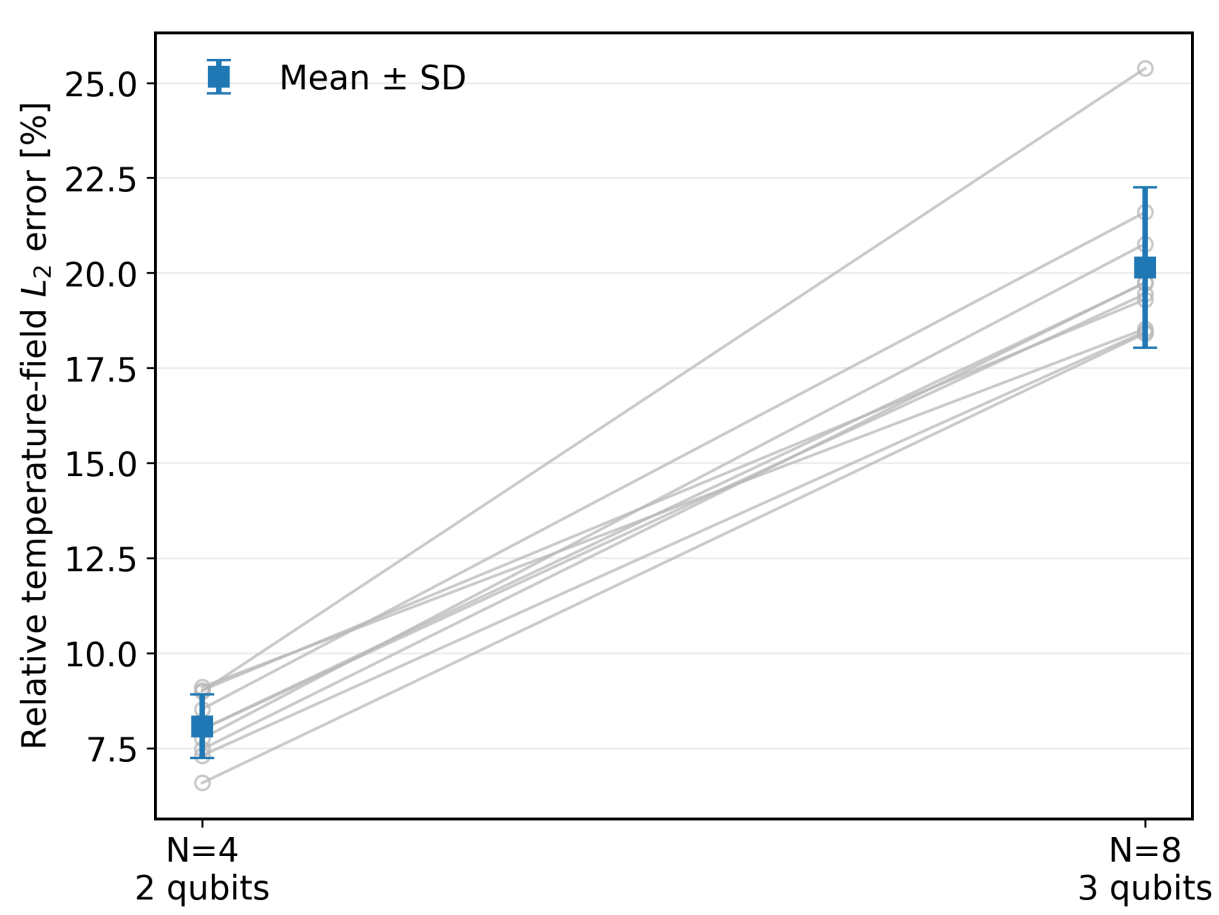

Figure 2. Paired relative temperature-field $L_2$ error for the confirmatory N=4 versus N=8 study (J=10). Each thin line joins matched observations from one independently scheduled hardware block; squares and error bars show mean ± one sample standard deviation. All ten paired differences are positive.

### 5.4 Secondary residual and Estimator-cost endpoints

The reconstructed linear-system residual shows a similar paired increase (Figure 3a). Mean relative residual is 6.47% for N=4 and 15.91% for N=8, yielding a paired increase of 9.44 percentage points (95% CI, 7.34-11.54; exact p=0.001953; Holm-adjusted p=0.003906). The resilience-0 Estimator VQLS cost is more variable (Figure 3b): mean cost increases from 0.0785 to 0.1250, with a paired difference of 0.0465 (95% CI, 0.0126-0.0804; exact and Holm-adjusted p=0.01953). Two blocks show negative cost differences even though every block shows a positive field-error difference, demonstrating that the variational cost and the recovered engineering field do not respond identically to hardware imperfections.

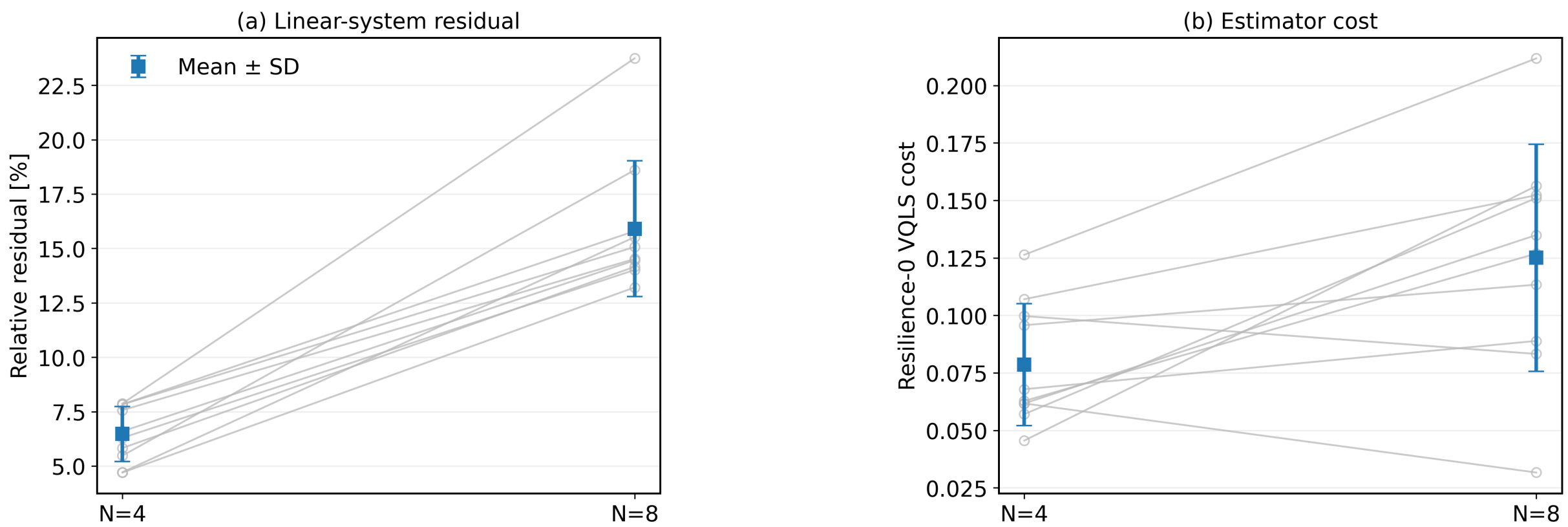

Figure 3. Secondary confirmatory hardware endpoints across J=10 paired blocks: (a) relative linear-system residual and (b) resilience-0 Estimator VQLS cost. Thin lines denote matched block-level observations; squares and error bars show mean ± one sample standard deviation.

Table 2. Confirmatory paired-block results (J=10). Differences are N=8 minus N=4.

| **Outcome** | **N=4 mean ± SD** | **N=8 mean ± SD** | **Paired difference** | **95% CI** | **Exact p** | **Holm p** |
|---|---|---|---|---|---|---|
| Relative $L_2$ error | 8.07 ± 0.84% | 20.14 ± 2.11% | +12.06 pp | 10.71 to 13.42 pp | 0.001953 | — |
| Relative residual | 6.47 ± 1.27% | 15.91 ± 3.12% | +9.44 pp | 7.34 to 11.54 pp | 0.001953 | 0.003906 |
| VQLS cost | 0.0785 ± 0.0265 | 0.1250 ± 0.0494 | +0.0465 | 0.0126 to 0.0804 | 0.01953 | 0.01953 |

*Note: pp = percentage points. Holm adjustment is applied only to the two pre-specified secondary endpoints.*

### 5.5 Exploratory N=16 scaling and readout extension

Across the five independently scheduled N=16 exploratory blocks, mean relative $L_2$ temperature-field error is 24.75±1.84%, mean relative residual is 24.72±3.17%, and mean resilience-0 VQLS cost is 0.1453±0.0571. Figure 4 places these descriptive values beside the confirmatory N=4 and N=8 summaries without treating the three configurations as a common inferential comparison. The N=16 field and residual are larger than the N=8 means, while the Estimator cost remains comparatively variable. This pattern is consistent with the nodewise profile in Figure 1: additional low-amplitude degrees of freedom substantially degrade full-field recovery even though the N=16 circuit depth is only modestly larger than that of N=8.

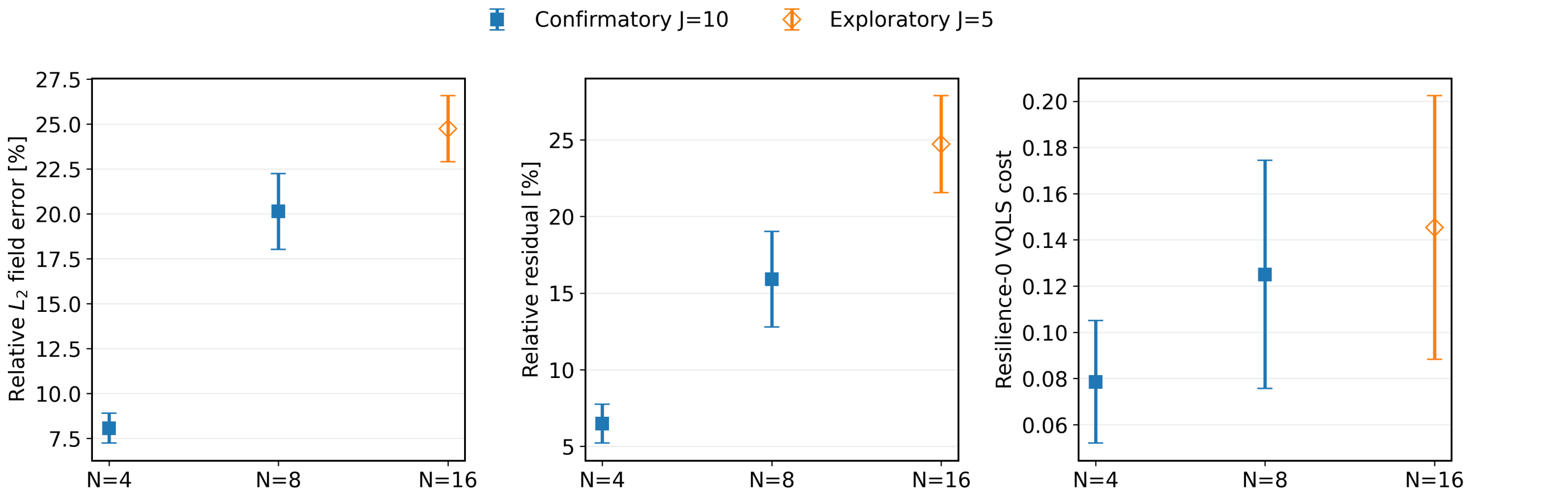

Figure 4. Descriptive progression of hardware metrics across N=4, N=8, and N=16: relative temperature-field $L_2$ error, relative residual, and resilience-0 VQLS cost. N=4 and N=8 are the J=10 confirmatory cases; N=16 is the separately acquired J=5 exploratory extension. Error bars denote one sample standard deviation. The figure is descriptive and does not imply a three-group inferential comparison.

### 5.6 N=16 shot-count sensitivity

The six-block follow-up directly tested whether additional Sampler shots recover the N=16 temperature field (Figure 5). Mean relative $L_2$ error is 23.64±1.27% at 2,048 shots, 24.57±1.21% at 8,192 shots, and 24.37±0.72% at 32,768 shots. The pre-specified 32,768-minus-2,048-shot paired difference is +0.74 percentage points (95% CI, -0.65 to 2.12; exact paired sign-flip p=0.281). Because a negative contrast denotes improvement, no reduction in mean field error is detected from the 16-fold shot increase. None of the six blocks shows a monotonic decrease in $L_2$ error across all three shot levels.

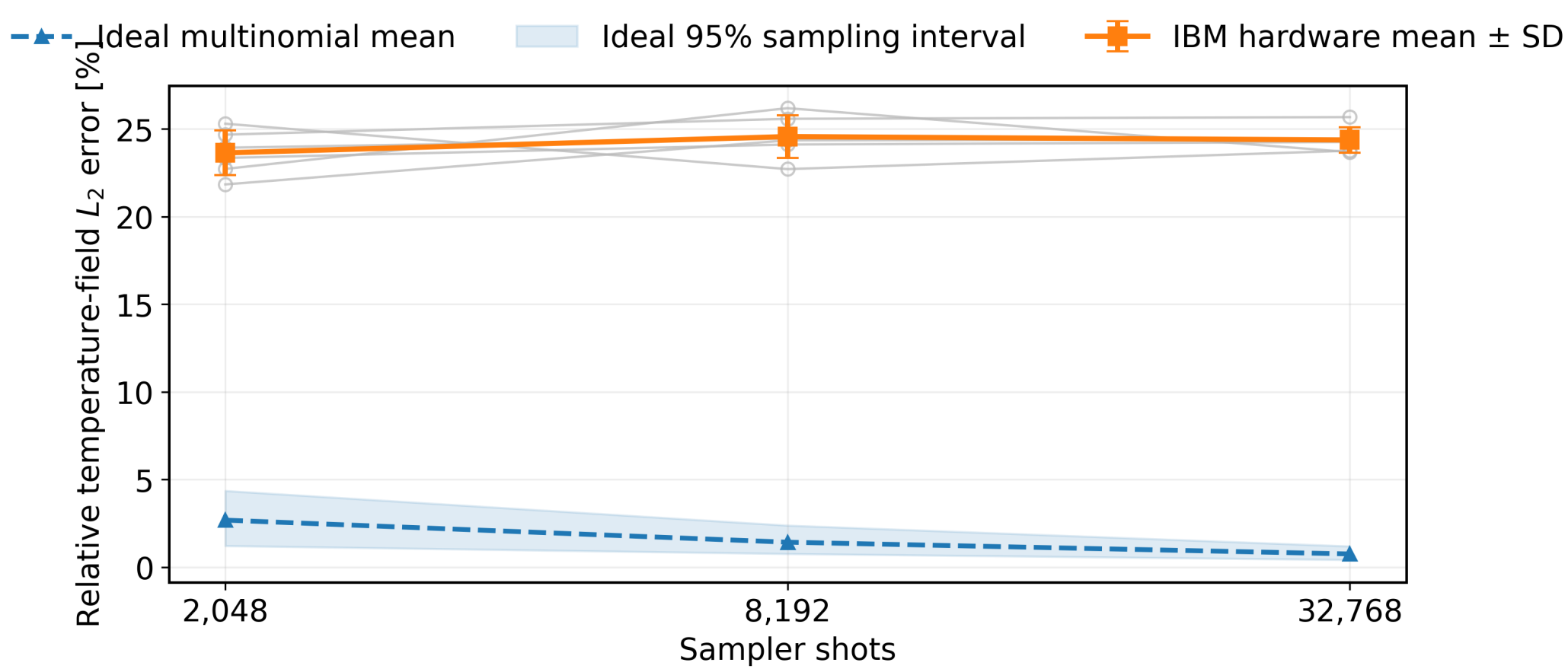

Figure 5. N=16 shot-count sensitivity of full-field temperature reconstruction. Thin lines connect the 2,048-, 8,192-, and 32,768-shot observations within each of six independently scheduled hardware blocks; squares and error bars show the IBM mean ± one sample standard deviation. The dashed reference shows the mean error predicted by zero-QPU ideal multinomial sampling, and the shaded region denotes its 95% model-sampling interval. The pre-specified 32,768-minus-2,048-shot hardware contrast is +0.736 percentage points (95% CI, -0.652 to 2.124; exact paired sign-flip p=0.281), indicating no detected improvement in mean full-field $L_2$ error from the 16-fold shot increase.

The hardware result contrasts sharply with the zero-QPU multinomial benchmark. Ideal sampling alone predicts mean field error of 2.68% at 2,048 shots, 1.42% at 8,192 shots, and 0.75% at 32,768 shots. Hardware error instead remains near 24% over the same range. Increasing shots nevertheless improves precision: the between-block standard deviation of hardware $L_2$ error falls from 1.27 to 0.72 percentage points, and the residual standard deviation falls from 1.86 to 0.39 percentage points. Mean residual decreases descriptively from 22.99% to 21.59%; the 32,768-minus-2,048-shot residual contrast is -1.41 percentage points (95% CI, -3.49 to 0.68; Holm-adjusted p=0.3125). Thus higher shot count reduces run-to-run dispersion, but no corresponding improvement in mean temperature-field accuracy is detected.

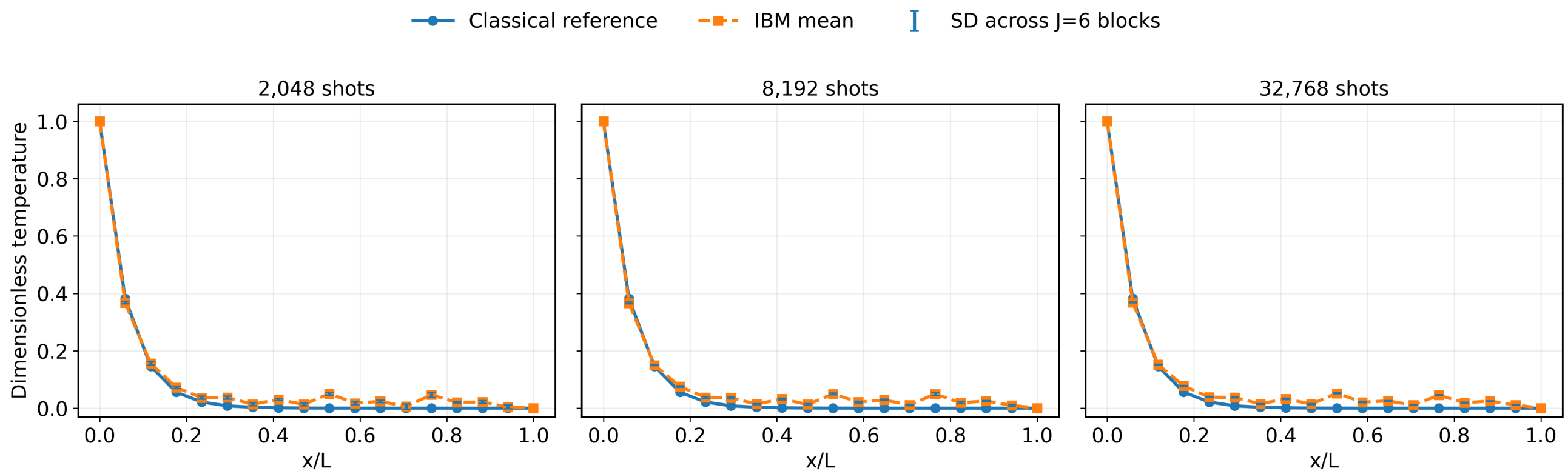

Figure 6. N=16 temperature profiles reconstructed at 2,048, 8,192, and 32,768 Sampler shots using the same locked circuit and physical layout. Hardware markers show the mean across J=6 independently scheduled shot-sensitivity blocks and error bars show one sample standard deviation at interior nodes. Higher shot counts reduce block-to-block variability, but the positive cold-side reconstruction floor persists relative to the classical reference.

### 5.7 QPU resource use

Charged QPU time was highly repeatable within each protocol. The ten-block N=4/N=8 confirmatory study consumed 620 QPU s in total (62 s per paired block). The five N=16 exploratory blocks consumed 265 QPU s (53 s per block). In the shot-count follow-up, every block consumed 10, 35, and 133 QPU s for the 2,048-, 8,192-, and 32,768-shot jobs, respectively, for 178 QPU s per block and 1,068 QPU s across six blocks. The three replicated datasets reported here therefore consumed 1,953 QPU s (32 min 33 s) in total.

## 6. Discussion

### 6.1 Thermal-field fidelity is a distinct requirement

The central heat-transfer result is that an essentially exact ideal VQLS state does not guarantee an accurate temperature field after execution and readout on current hardware. Across all three benchmark configurations, hot-side temperatures are reconstructed more faithfully than cold-side temperatures. The N=8 confirmatory profile develops a positive floor where the reference solution is small, and the exploratory N=16 profile extends that floor over a much larger fraction of the domain. This distinction matters because engineering use ultimately depends on the spatial field or on observables derived from it, not on the quantum cost alone. A boundary heat-flux estimate, an integrated stored-energy metric, and a local cold-side temperature can respond very differently to the same measured quantum state. Thermal quantities of interest should therefore be evaluated directly rather than inferred from a single global state-space metric.

### 6.2 Why the cold side is difficult to measure

The thermal decay is amplified by amplitude encoding because measurement probability is proportional to squared amplitude. The N=8 reference temperature falls from approximately 0.382 at the first interior node to $3.87\times10^{-4}$ at the final node; for N=16 the last interior value is approximately $1.75\times10^{-7}$. At 2,048 shots, several N=8 cold-side populations and twelve of the sixteen N=16 interior populations lie below one expected ideal count. Finite sampling therefore cannot reliably resolve those components, while device noise can populate states that ideally carry almost no probability. Under the positivity-based reconstruction used here, such populations become positive amplitudes and can produce a nonzero temperature floor after physical rescaling. The N=16 case makes this challenge especially visible because the circuit depth rises only modestly relative to N=8 while the number of thermally relevant low-amplitude components grows sharply.

### 6.3 Higher sampling improves precision, not mean field accuracy

The shot-count follow-up provides a controlled test of whether the N=16 field error is primarily an ordinary finite-sampling effect. Under ideal multinomial sampling, a 16-fold increase from 2,048 to 32,768 shots reduces mean $L_2$ error from 2.68% to 0.75%, approximately the expected square-root sampling trend. On hardware, the same increase changes the mean field error from 23.64% to 24.37%, with a paired 95% confidence interval spanning only -0.65 to +2.12 percentage points. At the same time, between-block variability decreases substantially. The combination of nearly unchanged mean error and reduced dispersion is inconsistent with finite-shot sampling as the dominant explanation for the observed N=16 field error under these conditions. Instead, it is consistent with persistent device-, circuit-, or readout-related errors that are repeatedly sampled more precisely as the number of shots increases.

This precision-versus-accuracy distinction is important for thermal simulation. More shots can stabilize an estimate without moving it closer to the physically correct field. The residual shows a modest descriptive decrease with shot count, whereas the full-field $L_2$ error does not, again showing that algebraic consistency and engineering-field fidelity need not vary together. For amplitude-encoded diffusion fields, simply increasing measurement count is therefore unlikely to be an efficient remedy once a hardware-induced population floor dominates low-amplitude basis states.

### 6.4 Benchmark complexity across N=4, N=8, and exploratory N=16

The experiment measures benchmark-configuration effects rather than a pure causal effect of N. Increasing from N=4 to N=8 changes logical qubits, ansatz parameters, circuit depth, two-qubit-gate count, Pauli-measurement workload, conditioning, and thermal-state dynamic range. The paired confirmatory design establishes a reproducible difference between those two locked configurations but does not isolate which complexity change is responsible. The exploratory N=16 extension introduces additional differences, including a four-qubit physical support acquired in a later hardware window. From N=8 to N=16, the condition number changes only from 4.35 to 4.80 and Sampler depth from 30 to 32, while two-qubit gates increase from 6 to 9 and H/G Pauli terms from 21/14 to 45/30. At the same time, the temperature-state dynamic range expands from roughly three to more than six orders of magnitude. The shot-count follow-up further shows that additional sampling alone does not recover the mean N=16 field. Taken together, the results support treating state preparation, hardware execution, full-field readout, and thermally meaningful observables as distinct layers of performance.

### 6.5 Implications for quantum heat-transfer simulation

For the present benchmark, classical solution is overwhelmingly cheaper than quantum execution; the value of the experiment is methodological rather than computational speed. The workflow carries a heat-transfer linear system from discretization through ideal verification to replicated hardware measurement while preserving a clear experimental unit and a physically interpretable output. The results also indicate that larger encoded dimension should not be pursued independently of the measurement strategy. For strongly nonuniform thermal fields, observable-focused approaches may be more informative than reconstructing every nodal temperature. Boundary heat rate, domain-averaged temperature, stored energy, selected sensor temperatures, or other low-dimensional functionals are natural targets if their measurement operators can be evaluated efficiently. Future work should extend to multidimensional conduction, variable coefficients, phase change, and conjugate thermal problems while retaining rigorous classical verification, independent hardware replication, and direct assessment of thermally relevant observables. Structured ansätze, preconditioning, mitigation, locality-aware formulations, and direct observable estimation may be more consequential than raw qubit count or shot count alone.

## 7. Limitations

The conclusions are intentionally narrow. Confirmatory inference concerns only the locked N=4 and N=8 configurations on one IBM backend and shared physical support. N=16 was added later as a separately locked, five-block exploratory extension on different physical qubits. The six-block shot-count follow-up was pre-specified within that later N=16 phase and balanced all shot-order permutations, but its small sample size yields discrete randomization p-values and does not constitute a formal equivalence test. The ideal multinomial control isolates finite-shot sampling only; it does not model gate errors, readout error, calibration drift, crosstalk, or other device effects. Shot conditions were separate Runtime jobs within each Batch, so within-block temporal differences cannot be eliminated completely. The positivity prior used for amplitude reconstruction is valid for this first-step diffusion problem but is not general to signed or complex systems. Variational parameters were optimized locally rather than with a QPU-in-the-loop optimizer, so the experiment isolates state execution and readout rather than end-to-end noisy optimization. Finally, the thermal physics is limited to one-dimensional constant-property conduction over one implicit step.

## 8. Conclusions

A variational quantum linear solver was applied to the sparse systems generated by an implicit transient heat-conduction step and evaluated on IBM ibm_miami. Classical direct solution and ideal VQLS simulation agree to approximately $10^{-8}$ relative error for N=4, N=8, and N=16, providing a controlled reference for hardware evaluation. In the pre-specified ten-block paired experiment, mean relative $L_2$ temperature-field error increases from 8.07% for N=4 to 20.14% for N=8, a paired increase of 12.06 percentage points (95% CI, 10.71-13.42; exact p=0.001953). A separately designated five-block N=16 extension yields 24.75±1.84% mean field error and exposes a persistent cold-side temperature floor across many low-amplitude components. A six-block shot-count follow-up then shows that increasing Sampler shots from 2,048 to 32,768 does not improve mean full-field accuracy: the paired difference is +0.74 percentage points (95% CI, -0.65 to 2.12; p=0.281), whereas ideal multinomial sampling predicts a reduction from 2.68% to 0.75%. Higher shot counts reduce between-block variability but leave the mean hardware field error essentially unchanged. Under the tested conditions, ordinary finite-shot sampling is therefore not the dominant limitation in N=16 full-field reconstruction. For heat-transfer applications, compact state encoding and reliable recovery of engineering quantities should be treated as separate requirements; observable-focused measurements of heat flux, stored energy, and selected temperatures may be more promising than exhaustive nodal readout.

## Data and Code Availability

The reproducibility package contains the heat-discretization and VQLS source code, statevector verification routines, immutable circuit locks, block-level N=4/N=8 confirmatory records, N=16 exploratory records, the complete shot-sensitivity records, IBM Runtime job identifiers, zero-QPU multinomial controls, statistical analysis code, nodewise temperature-profile summaries, and publication figures and tables. The source code used in this analysis will be made available by the authors upon reasonable request.